\documentclass[12pt]{article}

\usepackage{graphicx}
\usepackage{caption}
\usepackage{subcaption}
\usepackage{mathptmx}           % use Times fonts if available on your TeX system
\usepackage[applemac]{inputenc} % ansinew ou unicode in Linux
\usepackage{dsfont}
\usepackage{clrscode}             % Cormen & al. pseudo code package
\usepackage{url} 
\usepackage{color}
\usepackage[usenames,dvipsnames]{xcolor}
\usepackage{ marvosym }
\usepackage{enumitem}
\usepackage{natbib}
\usepackage{hyperref}
\usepackage[shortcuts]{extdash}

\begin{document}
\sloppy

\title{\bf Java Implementation of a Parameter-less Evolutionary Portfolio}

\author{   {\bf José C. Pereira}\\
            \small CENSE and DEEI-FCT\\
            \small Universidade do Algarve\\
            \small Campus de Gambelas\\
            \small 8005-139 Faro, Portugal\\
            \small unidadeimaginaria@gmail.com
\and
		{\bf Fernando G. Lobo}\\
            \small CENSE and DEEI-FCT\\
            \small Universidade do Algarve\\
            \small Campus de Gambelas\\
            \small 8005-139 Faro, Portugal\\
            \small fernando.lobo@gmail.com         
}
\date{}
\maketitle

\begin{abstract}
The Java implementation of a portfolio of parameter-less evolutionary algorithms is presented. The \emph{Parameter-less Evolutionary Portfolio} implements a heuristic that performs adaptive selection of parameter-less evolutionary algorithms in accordance with performance criteria that are measured during running time. At present time, the portfolio includes three parameter-less evolutionary algorithms: Parameter\=/less Univariate Marginal Distribution Algorithm, Parameter\=/less Extended Compact Genetic Algorithm, and Parameter\=/less Hierarchical Bayesian Optimization Algorithm. Initial experiments showed that the parameter-less portfolio can solve various classes of problems without the need for any prior parameter setting technique and with an increase in computational effort that can be considered acceptable.
\end{abstract}

%%%%%%%%%%%%%% INTRODUCTION %%%%%%%%%%%%%%%%%

\section{Introduction}\label{sec:intro}

The \emph{Parameter-less Evolutionary Portfolio} (P-EP) implements a heuristic that performs adaptive selection of parameter-less evolutionary algorithms (P-EAs) in accordance with performance criteria that are measured during running time. This heuristic is inspired by the parameter\=/less genetic algorithm (P-GA) \citep{HarikLobo:99} and was first proposed by \cite{Lobo:10a}. We direct the interested reader to these papers for a more general and detailed description of the heuristic itself. 

The main goal of this technical report is to present a Java implementation of the Parameter-less Evolutionary Portfolio (P-EPJava) whose source code is available for free download at \textcolor{Blue}{\href{https://github.com/JoseCPereira/2015ParameterlessEvolutionaryPortfolioJava}{https://github.com/JoseCPereira/2015ParameterlessEvolutionaryPortfolioJava}}.

The remainder of this paper is organized as follows. In Section \ref{sec:PGA} we briefly describe the main concept of the Parameter-less Evolutionary Portfolio. In Section \ref{sec:P-EAJava} we discuss the P-EPJava implementation itself and provide detailed instructions on how to use it and how to implement new problems with it.
%In Section \ref{sec:final} we make some final remarks about our current work related with PEBS and the P-EAJava.  

\section{The Parameter-less Evolutionary Algorithm}\label{sec:PGA}

The Parameter-less Evolutionary Portfolio always includes P-EAs that can be informally, but quantifiably, ordered by their increasing ``complexity''. The more complex P-EAs should be capable of tackling more difficult problems, usually at the expense of using an increased cost in model building. Following this order, the P-EP alternates between each algorithm on a continuous loop, giving the same amount of CPU time to all P-EAs in each iteration. Starting with a chosen initial time, $T_0$, the allowed CPU time is updated at each loop iteration to at least match the maximum time spent in one generation by any of the P-EAs. At the same time, because they are able to advance further in the search due to faster model building, simpler P-EAs are eliminated from the loop as soon as their current best average fitness is lower than the best average fitness of a more complex P-EA. 

The use of parameter\=/less algorithms allows P-EP to work as a black-box algorithm, without the need for any prior parameter settings. Additionally, P-EP is designed to run forever, physical constraints aside, because in practice the quality of the optimal solution is often unknown for many problems, making it impossible to distinguish, for instance, when an algorithm has reached the optimum result from when it simply got ``stuck'' in some plateau of the search space. Therefore, P-EP leaves to the user the decision when to stop the computation, based on the quality of the solutions already found and on the time and resources that she or he is willing to spend. 

%%%%%%%%%%%%%% P-EAJava P-EAJava %%%%%%%%%%%%%%%%%

\section{A Java Implementation of the Parameter-less Evolutionary Portfolio}\label{sec:P-EAJava}

At present time, the Java implementation of the Parameter-less Evolutionary Portfolio (P-EPJava) includes three P-EAs: Parameter\=/less Univariate Marginal Distribution Algorithm (P-UMDA)\footnote{See, for example, \cite{Muhlenbein:96} for a description of the original, non parameter-less version of the UMDA.}, Parameter\=/less Extended Compact Genetic Algorithm (P-ECGA) \citep{Lobo:00}, and Parameter\=/less Hierarchical Bayesian Optimization Algorithm (P-HBOA) \citep{PelikanHartLin:07}. As presented in detail in another arXiv report from the same authors, these three P-EAs (plus the P-GA) are also implemented in Java as independent algorithms. The source and binary files of those Java implementations are available for free download at \textcolor{Blue}{\href{https://github.com/JoseCPereira/2015ParameterlessEvolutionaryAlgorithmsJava}{https://github.com/JoseCPereira/2015ParameterlessEvolutionaryAlgorithmsJava}}.

All algorithms integrated in the evolutionary portfolio must be parameter-less in the sense that they use the population sizing method employed by the Parameter-less Genetic Algorithm \citep{HarikLobo:99}. In addition, each P-EA must satisfy the following two constraints in order to be integrated in the P-EPJava:

\begin{enumerate}
\item The algorithm represents possible solutions (individuals) as strings of zeros and ones.
\item All individuals have the same string size.
\end{enumerate}
 
  \subsection{How to use the P-EPJava}
 
  The P-EPJava is a Java application developed with the Eclipse\footnote{Version: Kepler Service Release 2} IDE. The available code is already compiled and can be executed using the command line.
 
 \paragraph{Run the P-EAJava from a command line}
 
\begin{enumerate}
\item Unzip the source file \textit{2015ParameterlessPortfolio.zip} to any directory.
\item Open your favourite terminal and execute the command\\ 
\vspace{-.5cm}

{\tt{~~cd [yourDirectory]/2015ParameterlessEvolutionaryPortfolio/bin}}\\
\vspace{-.3cm}

where {\tt{[yourDirectory]}} is the name of the directory chosen in step 1.
 
\item Execute the command\\ 
\vspace{-.5cm}

{\tt{~~java com/z\_PORTFOLIO/PORTFOLIO ./PortParameters.txt}}
\end{enumerate}

The argument ``PortParameters.txt'' is in fact the name of the file containing all the options concerning the portfolio settings and can be changed at will.

After each execution of a single or multiple runs, the P-EPJava produces one output file --\emph{PORTFOLIO\_*\_*.txt} -- that records how each run progressed in terms of time allowed to each algorithm, which algorithms were deactivated and when, number of fitness calls performed by each algorithm, best individual and average fitnesses, the evolution of the population sizes, among other relevant information. All this information is also displayed on the screen during execution time.

At present time, the P-EPJava version made available with this paper already includes the following set of test problems:\\

\textit{ZERO} Problems \textit{\qquad \qquad \qquad \qquad \qquad ONE} Problems
\vspace{.4cm} 

\begin{tabular}{rlrl}
~~0 $\rightarrow$ & ZeroMax & ~~10 $\rightarrow$ & OneMax\\
~~1 $\rightarrow$ & Zero Quadratic & ~~11 $\rightarrow$ & Quadratic\\
~~2 $\rightarrow$ & Zero 3-Deceptive & ~~12 $\rightarrow$ & 3-Deceptive\\
~~3 $\rightarrow$ & Zero 3-Deceptive Bipolar & ~~13 $\rightarrow$ & 3-Deceptive Bipolar\\
~~4 $\rightarrow$ &  Zero 3-Deceptive Overlapping & ~~14 $\rightarrow$ & 3-Deceptive Overlapping\\
~~5 $\rightarrow$ & Zero Concatenated Trap-k & ~~15 $\rightarrow$ & Concatenated Trap-k\\
~~6 $\rightarrow$ & Zero Uniform 6-Blocks & ~~16 $\rightarrow$ & Uniform 6-Blocks
\end{tabular}\\ 

\

\textit{Hierarchical} Problems	
\vspace{.4cm} 

\begin{tabular}{rl}
~~21 $\rightarrow$ &  Hierarchical Trap	One\\
~~22 $\rightarrow$ & Hierarchical Trap	Two
\end{tabular}\\

\

The Zero problems always have the string with all zeros as their best individual. The One problems are the same as the Zero problems but their best individual is now the string with all ones. A description of these problems can be found, for instance, in \cite{PelikanPazGoldberg:2000}. The Hierarchical problems are thoroughly described in \cite{Pelikan:05}. 

It is also possible to define a noisy version for any of the previous problems. This is done by adding a non-zero Gaussian noise term to the fitness function.

The source code that implements all the problems mentioned in this section can be found in the file  \textit{src/com/z\_PORTFOLIO/Problem.java}.

As mentioned previously, all options concerning the evolutionary portfolio are in the file \textit{PortParameters.txt}. In particular, it is in this file that are made the choices for the problem to be solved.

To choose a particular problem the user must set the value of the following three options:\\

\begin{tabular}{rl}
~~Line 81:  &  \textit{problemType}\\
~~Line 90:  & \textit{stringSize}\\
~~Line 107: & \textit{sigmaK} ~~~~~(defines the noise component)
\end{tabular}\\

%\newpage
All other options are set to default values and their role in the behaviour or the portfolio is explained with detail in the file's comments. This is also true for the parameters specific to the parameter-less strategy and to each of the implemented parameter-less algorithms which are defined in four separate files:\\

\begin{tabular}{rl}
PARAMETER-LESS: &ParParameters.txt\\
UMDA:  &UMDAParameters.txt\\
ECGA:  &ECGAParameters.txt\\
HBOA:  &HBOAParameters.txt\\	
\end{tabular}\\

Note that the default settings defined in these four files were chosen to ensure a robust behavior of the corresponding algorithms, in accordance with current theory. Therefore, the user is advised to proceed with caution when performing any changes in those settings. In fact, the whole idea behind the portfolio and the parameter-less strategy is to eliminate the need of such fine tuning when solving a particular problem. After choosing a problem to be solved and a particular algorithm to solve it, the user has only to press the start button and wait until the P-EPJava finds a solution with good enough quality.

\begin{figure}[t!]
\centering
\includegraphics[width=0.85\textwidth]{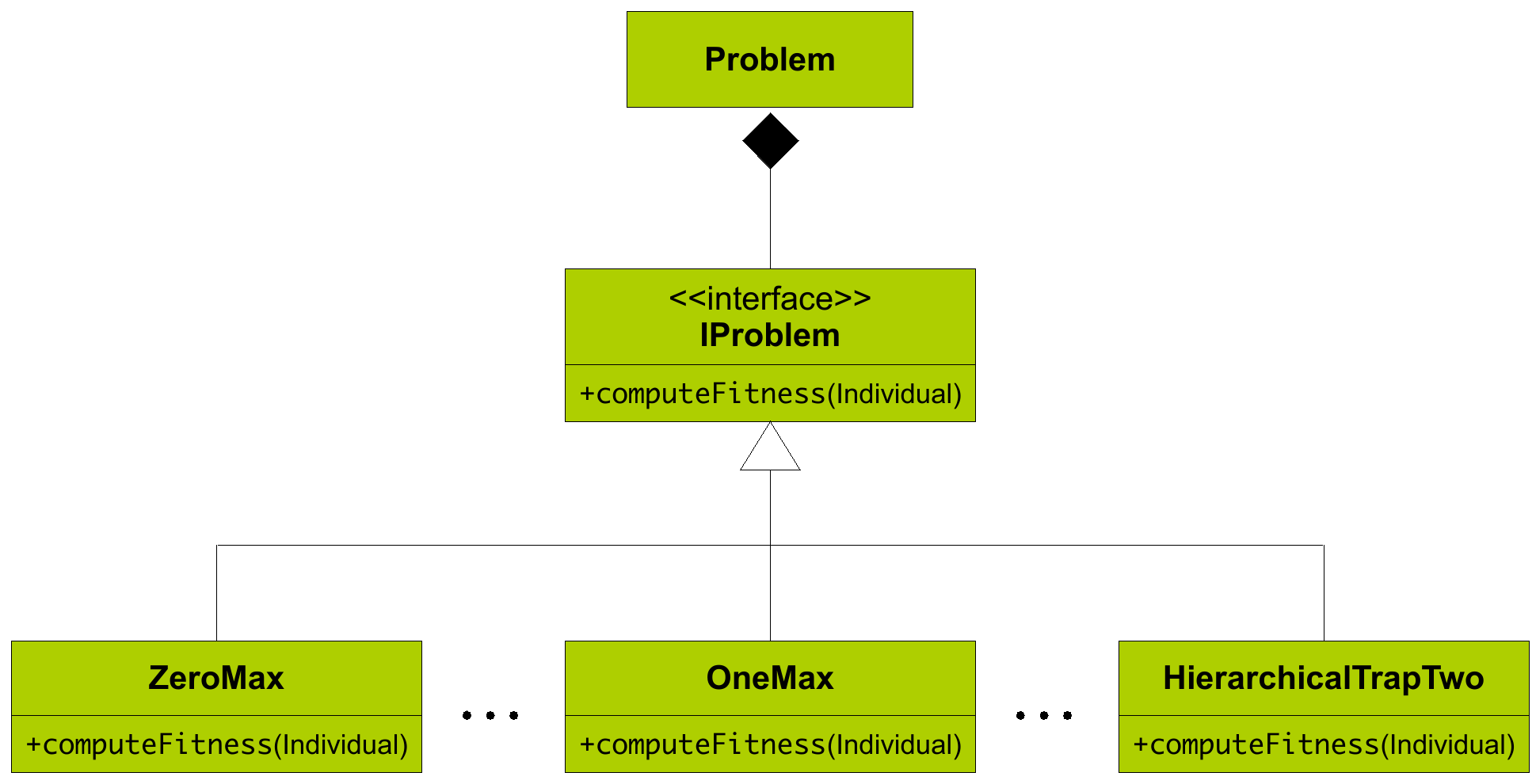}
\caption{The P-EPJava uses the design pattern strategy \citep{Gamma:95} to allow an easy implementation of new problems to be solved by the framework.}
\label{fig:IProblem}
\end{figure}

\subsection{How to implement a new problem with P-EAJava}

The P-EPJava uses the design pattern strategy \citep{Gamma:95} to decouple the implementation of a particular problem from the remaining portfolio structure (see Figure \ref{fig:IProblem}). As a consequence, to plug in a new problem to the framework it is only necessary to define one class that implements the interface IProblem and change some input options to include the new choice. The interface IProblem can be found in the file \textit{src/com/z\_PORTFOLIO/Problem.java}.

In the following let us consider that we want to solve a new problem called \textit{NewProblem} with one of the parameter-less algorithms. To plug in this problem it is necessary to:

\newpage
\begin{enumerate}
\item Define a class called \textit{NewProblem} in the file \textit{src/com/z\_PORTFOLIO/Problem.java}. The signature of the class will be
\begin{figure}[h!]
\centering
\includegraphics[width=0.65\textwidth]{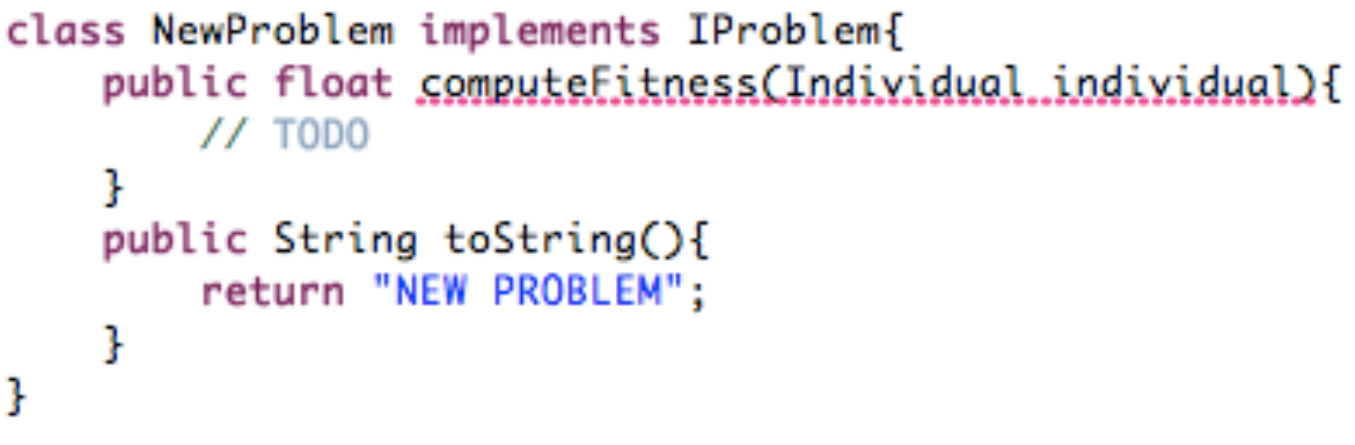}
\end{figure}
\vspace{-.5cm}
\item Code the body of the function computeFitness(Individual) according to the nature of problem \textit{newProblem}. The class \textit{Individual} provides all the necessary functionalities to operate with the string of zeros and ones that represents an individual  (e.g., getAllele(int)). This class can be found in the file \textit{src/com/z\_PORTFOLIO/Individual.java}.

\item To define the new problem option, add the line

\includegraphics[width=.95\textwidth]{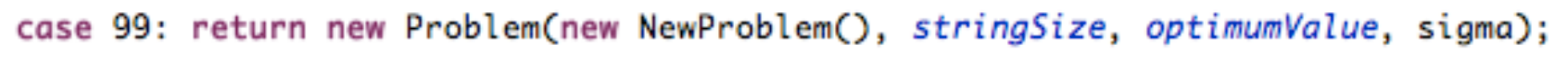}
 to the \textbf{switch} command in line 174 of the file \textit{src/com/z\_PORTFOLIO/PortParameter.java}. The case number -- 99 -- is a mere identifier of the new problem option. The user is free to choose other value for this purpose. The rest of the line is to be written verbatim. 
 
\item Validate the new problem -- option value 99 -- by adding the case \textit{problemType} == 99 to the conditional in line 111 of the same \textit{PortParameter.java} file.

\end{enumerate}

\vspace{.2cm}
Although not strictly necessary, it is also advisable to keep updated the problem menu in the file \textit{PortParameters.txt}.

\section*{Acknowledgements}
The current version of the P-EAJava is one of the by-products of the activities performed by the PhD fellow, José C. Pereira within the doctoral program entitled \textit{Search and Optimization with Adaptive Selection of Evolutionary Algorithms}. The work program is supported by the Portuguese Foundation for Science and Technology with the doctoral scholarship SFRH/BD/78382/2011 and with the research project PTDC/EEI-AUT/2844/2012.

%%%%%%%%%%%%%%%%%%%%%%%%%%%%% Main Text End %%%%%%%%%%%%%%%%%%%%%%%%%%%%%%%%%

%\bibliographystyle{apalike}
%\bibliographystyle{plainnat}
%\bibliography{PhDReferences}

\end{document}